\documentclass{jetpl}
\twocolumn
\title{Fermions with cubic and quartic spectrum}

\rtitle{Fermions with cubic and quartic spectrum}

\sodtitle{Fermions with cubic and quartic spectrum}

\author{Tero T. Heikkil\"a $^{*}$\/\thanks{e-mail: tero.heikkila@tkk.fi}
and G.E. Volovik $^{*+}$
\/\thanks{e-mail: volovik@boojum.hut.fi}
}

\rauthor{T.T. Heikkil\"a, G.E.Volovik}

\sodauthor{Heikkil\"a, Volovik}

\address{
$^{*}$ Low Temperature Laboratory, Aalto University, School of Science and
Technology, P.O. Box 15100, FI-00076 AALTO, Finland
\\
$^+$ Landau Institute for Theoretical Physics RAS, Kosygina 2,
119334 Moscow, Russia
  }

\dates{October, 2010; preprint arXiv:1010.0393}{*}


\abstract{We study exotic fermions with spectrum $E^2 \propto p^{2N}$. Such
  spectrum emerges in the vicinity of the Fermi point with multiple
  topological charge $N$, if special symmetry is obeyed. When this
  symmetry is violated, the multiple Fermi point typically splits into
  $N$ elementary Fermi points -- Dirac points with $N=1$ and spectrum
  $E^2 \propto p^{2}$. }

\begin{document}

\maketitle

\newcommand  {\version}{v15}

\section{Introduction}\label{sec:introduction}

 There is a fundamental interplay of symmetry and topology in physics, both in condensed matter and relativistic quantum fields. Traditionally the first role was played by symmetry (symmetry classification of crystals, liquid crystals, magnets, superconductors, superfluids, etc.).  The phenomenon of spontaneously broken symmetry remains one of the major tools in physics.  The last decades demonstrated the opposite tendency in which topology is primary. Topology in momentum space is becoming the main characteristics of  quantum vacua --  ground states of the system at   $T=0$
\cite{NielsenNinomiya1981,VolovikMineev1982,AvronSeilerSimon1983,Semenoff1984,NiuThoulessWu1985,Haldane1988,
Volovik1988,Volovik2003,HasanKane2010,Kitaev2009,SCZhang2009b,Schnyder2008,Stone2010,Varney2010}. For example in topological matter, nodes in the spectrum of fermionic quasiparticles are protected by topology, and this property is insensitive to the details of the microscopic many-body Hamiltonian
 (see e.g. review  \cite{Volovik2003}).
The momentum-space topological invariants determine universality classes of the topological matter and the type of the effective theory which  emerges at low energy and low temperature. In many cases they also give rise to emergent symmetry. Examples are provided by the nodes of the energy spectrum characterized by the elementary topological charges, $N=+1$ or $N=-1$. Close to such nodes the effective Lorentz invariance emerges: the Fermi point with $N=+1$ or $N=-1$ represents the Dirac point and the spectrum forms the relativistic Dirac cone. This is the consequence of the Atiyah-Bott-Shapiro construction  \cite{Horava2005}.

However, in many systems (including condensed matter and relativistic quantum vacua), the Fermi points with $N=+1$ or $N=-1$ may merge together forming the points either with $N=0$ or with multiple $N$ (i.e. $|N| >1$ \cite{VolovikKonyshev1988}). In this case topology and symmetry become equally important, because it is the symmetry which may stabilize the degenerate node.
Example is provided by the Standard Model of particle physics, where 16 fermions of one generation have degenerate Dirac point at ${\bf p}=0$ with the trivial total topological charge $N=8-8=0$. In the symmetric phase of  Standard Model the nodes in the spectrum survive due to a discrete symmetry between the fermions and they disappear in the non-symmetric phase forming the fully gapped vacuum \cite{Volovik2003}. In case of degenerate Fermi point with $|N| >1$, situation is more diverse. Depending on symmetry, interaction between fermionic flavors may lead to splitting of the multiple Fermi point to elementary Dirac points \cite{KlinkhamerVolovik2005}; or gives rise to the essentially non-relativistic energy spectrum $E_\pm(p\rightarrow 0) \rightarrow \pm p^N$, which corresponds to different scaling for space and time in the infrared: ${\bf r}\rightarrow \lambda {\bf r}$, $t\rightarrow \lambda^N t$. The particular case of anisotropic scaling with  $N=3$ was suggested by Horava for quantum gravity at short distances \cite{HoravaPRL2009,HoravaPRD2009,Horava2008}, and anisotropic scaling in the infrared in Ref. \cite{Horava2010}. 

The non-linear spectrum arising near the Fermi point with $N=2$ has been discussed for different systems including graphene, double cuprate layer in high-$T_c$ superconductors, surface states of topological insulators and neutrino physics
 \cite{Volovik2001,Volovik2003,Volovik2007,Manes2007,Dietl-Piechon-Montambaux2008,Chong2008,Banerjee2009,Sun2010,Fu2010}. 
The spectrum of (quasi)particles in the vicinity of the doubly degenerate node (say, with topological charge $N\equiv N_3=\pm 2$ for Fermi points of co-dimension 3) depends on symmetry.     One may obtain:  two Weyl fermions, if there is some special symmetry;
exotic massless fermions with quadratic dispersion at low energy, 
\begin{equation}
E_\pm({\bf p})\approx \pm \frac{p^2}{2m} \,;
\label{WeylQuadratic}
\end{equation}
or   semi-Dirac fermions with linear dispersion in one direction and quadratic dispersion in the other, 
  \begin{equation}
E_\pm({\bf p})\approx \pm \sqrt{c^2p_z^2 + \left(\frac{p_\perp^2}{2m}\right)^2 } \,.
\label{SemiDirac}
\end{equation}

Here we discuss the class  of effective Hamiltonians which has a Fermi point with higher degeneracy, described by the symmetry protected topological invariant $N=3$ and $N=4$, and  quantum phase transitions which may occur in these systems where topology of the spectrum changes.

\section{Cubic spectrum}\label{sec:cubic}

Let us consider first the case with $N=3$. Examples are three families of right-handed Weyl 2-component fermions in particle physics; three cuprate layers in high-$T_c$ superconductors; three graphene layers, etc. If the Fermi point is topologically protected, i.e. there is a conserved topological  invariant $N$, the node in the spectrum cannot disappear even in the presence of interaction, but it can split into $N$ nodes with elementary charge $N=1$. The splitting can be prevented if there is a symmetry in play, such as rotational symmetry. Here we provide an example of such symmetry.

For simplicity we study the 2+1 systems.  In general the nodal points in 2+1 dimensions (Fermi points of co-dimension 2) obey $Z_2$ topology  \cite{Horava2005,Volovik2007}, that is why we also need an additional symmetry $K$ which extends the group $Z_2$ to the full group $Z$ and thus makes the multiple Fermi point possible. The corresponding  invariant protected by symmetry $K$
\cite{Volovik2007,Beri2009} is:
\begin{equation}
N^K= \frac{1}{4 \pi i}~
{\bf tr}\oint_C   dl 
~ K G(\omega=0,{\bf p})\partial_l G^{-1}(\omega=0,{\bf p}) \,,
\label{MasslessTopInvariant1+1}
\end{equation}
where $G$ is  Green's function matrix, and $C$ is contour around the Fermi point in 2D momentum space $(p_x,p_y)$, or around the Fermi surface if the Fermi point expands to the Fermi surface. The matrix $K$  commutes or anti-commutes with matrix $G(\omega=0,{\bf p})$. The single particle Green's function at zero energy represents the effective single-particle  Hamiltonian ${\cal H}({\bf p})=G^{-1}(\omega=0,{\bf p})$. 
 
 \begin{figure}[h]
\centering
\includegraphics[width=8cm]{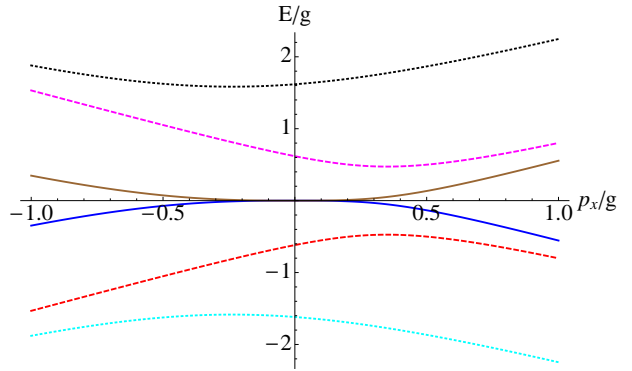}
\caption{\textbf{Fig.~1~}
Spectrum of the Hamiltonian \eqref{3x3a} showing cubic
  dispersion for the lowest two eigenvalues around the point
  $\mathbf{p}=0$. Different colors correspond to different
  eigenvalues, spectrum is shown as function of $p_x$ at $p_y=0$. The spectra have been calculated with equal
  coupling strengths $g_{12}=g_{13}=g_{23}=g$.}
  \label{Cubic_spectrum}
\end{figure}

We consider 3 species (families or flavors) of fermions, each of them being described by the invariant $N\equiv N^K=+1$ and an effective relativistic  Hamiltonian  emerging in the vicinity of the Fermi point
\begin{equation}
{\cal H}_0({\bf p})=\boldsymbol{\sigma}\cdot{\bf p}= \sigma_x p_x + \sigma_y p_y \,.
\label{NonInteractingSpecies}
\end{equation}
The matrix $K=\sigma_z$ anticommutes with the Hamiltonian. This supports 
the topologically protected node in spectrum, which is robust to interactions. The position of the node here is chosen at ${\bf p}=0$:
\begin{equation}
E^2=p^2 \,.
\label{NonInteractingSpectrum}
\end{equation}
The total topological charge of three nodes at ${\bf p}=0$ of three fermionic species is $N^K=+3$.
Let us now introduce mixing of the fermions, which obeys some symmetries which may follow from the underlying microscopic theory. We consider two cases
\begin{equation}
{\cal H}_1({\bf p})=
\left( \begin{array}{ccc}
\boldsymbol{\sigma}\cdot{\bf p} &g_{12}\sigma^+&g_{13}\sigma^-\\
g_{21}\sigma^-&\boldsymbol{\sigma}\cdot{\bf p} &g_{23}\sigma^+\\
g_{31}\sigma^+&g_{32}\sigma^-&\boldsymbol{\sigma}\cdot{\bf p}
\end{array} \right)
\,.
\label{3x3b}
\end{equation}
and
\begin{equation}
{\cal H}_2({\bf p})=
\left( \begin{array}{ccc}
\boldsymbol{\sigma}\cdot{\bf p} &g_{12}\sigma^+&g_{13}\sigma^+\\
g_{21}\sigma^-&\boldsymbol{\sigma}\cdot{\bf p} &g_{23}\sigma^+\\
g_{31}\sigma^-&g_{32}\sigma^-&\boldsymbol{\sigma}\cdot{\bf p}
\end{array} \right)
\,,
\label{3x3a}
\end{equation}
where $\sigma^\pm=\frac{1}{2}(\sigma_x\pm i\sigma_y)$ are ladder operators.
Both Hamiltonians  anti-commute with $K=\sigma_z$ and thus mixing preserves the topological charge
$N^K$ in \eqref{MasslessTopInvariant1+1}. 
Hamiltonian \eqref{3x3b} is symmetric with respect to the $Z_3$ group of permutations if all the couplings $g_{mn}$ are equal. For general couplings,
Hamiltonian \eqref{3x3b} is symmetric under the $Z_3$ group of rotations by $2\pi/3$ combined with gauge transformations from the $SU(3)$ family group: 
\begin{equation}
 {\cal H}_1({\hat O}_{\rm orb} {\bf p})={\hat U}^+_{\rm gauge} {\hat U}^+_{\rm spin} {\cal H}_1( {\bf p}){\hat U}_{\rm spin}{\hat U}_{\rm gauge} \,.
 \label{3-fold_symmetry}
\end{equation}
 Here ${\hat O}_{\rm orb}$ and ${\hat U}_{\rm spin}$ are orbital rotations by $2\pi/3$ in momentum and spin space respectively, and ${\hat U}_{\rm gauge}={\rm diag}\left(1, e^{2\pi i/3},  e^{-2\pi i/3}\right)$ is the element of the $SU(3)$ family group. 

\begin{figure}[h]
\centering
\includegraphics[width=8cm]{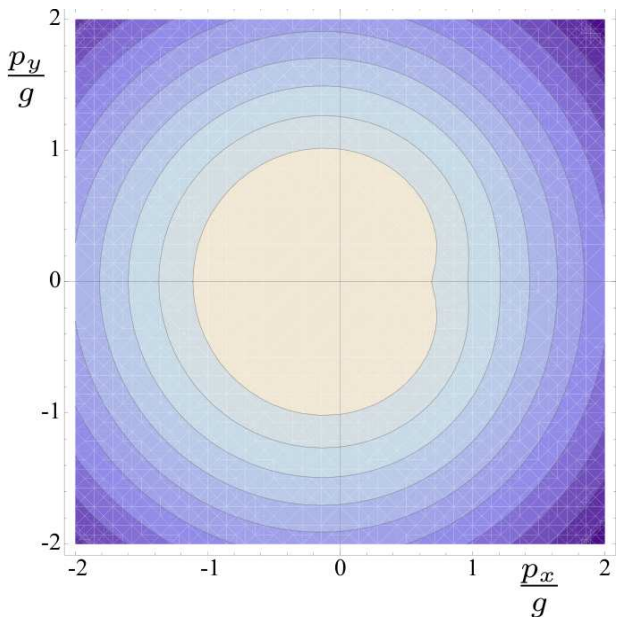}
\caption{\textbf{Fig.~2~}
Contour plot showing the equi-energy curves in the vicinity of the degenerate Fermi point with $N_K=3$ and cubic dispersion. Here $g_{12}=g_{23}=g=g_{13}/2$.}
  \label{cubic_dispersion_contour}
\end{figure}

The Hamiltonian \eqref{3x3a}  at $p_x=p_y=0$ is independent of the spin rotations up to a global phase of the coupling constants. Under spin rotation by angle $\theta$ all elements in the upper triangular matrix are multiplied by  $e^{i\theta}$, while all elements in the lower triangular matrix are multiplied by  $e^{-i\theta}$. This symmetry of triangular matrices generates specific property of the spectrum. 
Mixing between fermions does not split the multiple Fermi point at $p=0$, as a result the gapless branch of spectrum in Fig. \ref{Cubic_spectrum}  has the cubic form at low energy, $E\rightarrow 0$, which corresponds to the topological charge $N^K=+3$:
\begin{equation}
E^2\approx \gamma_3^2 p^6~~,~~ \gamma_3=\frac{1}{|g_{12}||g_{23}|}\,.
 \label{CubicSpectrum}
\end{equation}
In the low-energy limit the spectrum in the vicinity of the multiple Fermi point \eqref{CubicSpectrum} is symmetric under rotations. But in general the spectrum is not symmetric  as demonstrated in Figs. \ref{Cubic_spectrum} and \ref{cubic_dispersion_contour}. There is only the symmetry with respect to reflection, $(p_x,p_y) \rightarrow (p_x,-p_y)$. The rotational symmetry of spectrum \eqref{CubicSpectrum} is an emergent phenomenon.
 For trilayer graphene this spectrum has been discussed in Ref. \cite{Guinea2006}.

The cubic spectrum \eqref{CubicSpectrum} becomes singular when either one of the coupling constants $g_{12}$ or $g_{23}$ nullify.
At this point of topological quantum phase transition, the gapless spectrum has  two branches: one with quadratic dispersion $E_\pm=\pm p^2/2m$ and another one with linear
dispersion $E_\pm=\pm p$. 
Note that in the limit $p\rightarrow 0$ the spectrum \eqref{CubicSpectrum} does not depend on $g_{31}$, i.e. the spectrum is not symmetric under permutations. This is the property of the Hamiltonian ${\cal H}_2({\bf p})$: there is no permutation symmetry even if the couplings are the same, $g_{12}=g_{23}=g_{31}$.

\begin{figure}[h]
\centering
\includegraphics[width=8cm]{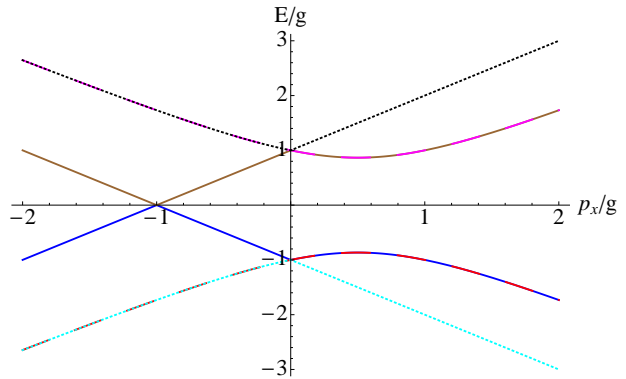}
\caption{\textbf{Fig.~3~}
Spectrum of the permutation symmetric Hamiltonian \eqref{3x3b} shows three Dirac points (Fermi points with $N=1$ and linear dispersion), located at $\{p_x,p_y\}/g=\{-1,0\},\{1/2,\sqrt{3}/2\},\{1/2,-\sqrt{3}/2\}$, first of them is shown in the figure. Different colors correspond to different
  eigenvalues, spectrum is shown as function of $p_x$ at $p_y=0$. The spectra have been calculated with equal coupling strengths $g_{12}=g_{13}=g_{23}=g$ in \eqref{6x6}.}
\label{permuting_hamiltonian_spectrum}
\end{figure}

\begin{figure}[h]
\centering
\includegraphics[width=8cm]{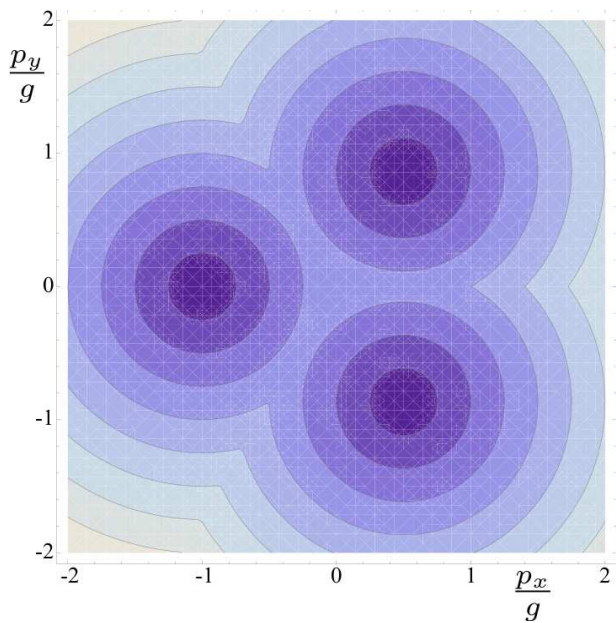}
\caption{\textbf{Fig.~4~}
Splitting of Fermi point with $N=3$ into three Fermi points with $N=1$ each in the spectrum of the Hamiltonian \eqref{3x3b}. We plotted the case of equal
  coupling strengths $g_{12}=g_{13}=g_{23}=g$, corresponding to full permutation symmetry.}
  \label{permuting_hamiltonian_spectrum_contour}
\end{figure}

At $g_{31}=0$ one also has the quantum phase transition, since at $g_{31}=0$ the Hamiltonians \eqref{3x3a} and \eqref{3x3b} coincide: ${\cal H}_2({\bf p},g_{31}=0)={\cal H}_1({\bf p},g_{31}=0)$. Then when $g_{31}$ grows from zero in \eqref{3x3b} the multiple Fermi point splits into 3 elementary Dirac points, i.e. Fermi points  with $N^K=+1$. 
For equal coupling strengths $g_{12}=g_{13}=g_{23}=g$ one has
\begin{equation}
{\cal H}_1({\bf p})=
\left( \begin{array}{cccccc}
0&p_+&0&g&0&0\\
p_-&0&0&0&g&0\\
0&0&0&p_+&0&g\\
g&0&p_-&0&0&0\\
0&g&0&0&0&p_+\\
0&0&g&0&p_-&0
\end{array} \right)
\,,
\label{6x6}
\end{equation}
where $p_\pm=p_x\pm ip_y$. In this case the full permutation symmetry takes place and three Fermi points with $N=1$ each  form the configuration obeying the three-fold rotational symmetry, see  Figs. \ref{permuting_hamiltonian_spectrum}
 and \ref{permuting_hamiltonian_spectrum_contour}. 
For general couplings, zeroes are located at $(p_x+ip_y)^3=-g_{12}g_{23}g_{31}$, also obeying the $Z_3$ symmetry in \eqref{3-fold_symmetry}.
As one of the couplings vanishes, the Fermi points merge at $p=0$ and produce either a linear, quadratic or cubic dispersion relation, depending on which of the couplings vanish.

\section{Fermi points with higher degeneracy}

\begin{figure}[h]
\centering
\includegraphics[width=8cm]{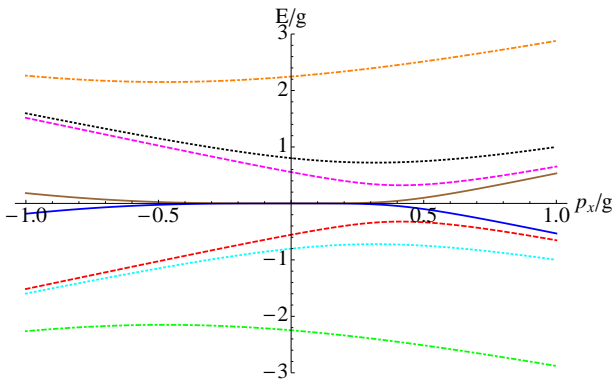}
\caption{\textbf{Fig.~5~}
Spectrum of the Hamiltonian \eqref{4x4a} showing quartic
  dispersion for the lowest two eigenvalues around the point
  $\mathbf{p}=0$. Different colors correspond to different
  eigenvalues, spectrum is shown as function of $p_x$ at $p_y=0$. The spectra have been calculated with equal coupling strengths $g_{12}=g_{13}=g_{23}=g_{14}=g_{24}=g_{34}=g$.}
\end{figure}

In case of four fermionic species, the mixing which does not produce splitting of the Fermi point is obtained by the same principle as in \eqref{3x3a}: all matrix elements above the main diagonal contain only  $\sigma^+$
(or $\sigma^-$):
\begin{equation}
{\cal H}_2({\bf p})=
\left( \begin{array}{cccc}
\boldsymbol{\sigma}\cdot{\bf p} &g_{12}\sigma^+&g_{13}\sigma^+&g_{14}\sigma^+\\
g_{21}\sigma^-&\boldsymbol{\sigma}\cdot{\bf p} &g_{23}\sigma^+&g_{24}\sigma^+\\
g_{31}\sigma^-&g_{32}\sigma^-&\boldsymbol{\sigma}\cdot{\bf p}&g_{34}\sigma^+\\
g_{41}\sigma^-&g_{42}\sigma^-&g_{42}\sigma^-&\boldsymbol{\sigma}\cdot{\bf p}
\end{array} \right)
\,.
\label{4x4a}
\end{equation}
Then again under spin rotation by angle $\theta$ all elements in the upper triangular matrix are multiplied by  $e^{i\theta}$, while all elements in the lower triangular matrix are multiplied by  $e^{-i\theta}$. As a result the multiple Fermi point with $N=4$ is preserved giving rise to the quartic spectrum in vicinity of the Fermi point:
\begin{equation}
E^2=\gamma_4^2 p^8~~,~~ \gamma_4=\frac{1}{|g_{12}||g_{23}||g_{34}|}\,.
 \label{QuarticSpectrum}
\end{equation}
For the tetralayer graphene this spectrum was suggested in Ref. \cite{Mak2010}.
Again the rotational symmetry emerges new the Fermi point, but spectrum is not symmetric under permutations. In fact, there is no such $4\times 4$ matrix that would consist of couplings described by ladder operators and which would be symmetric under permutations.

\begin{figure}[h]
\centering
\includegraphics[width=7.7cm]{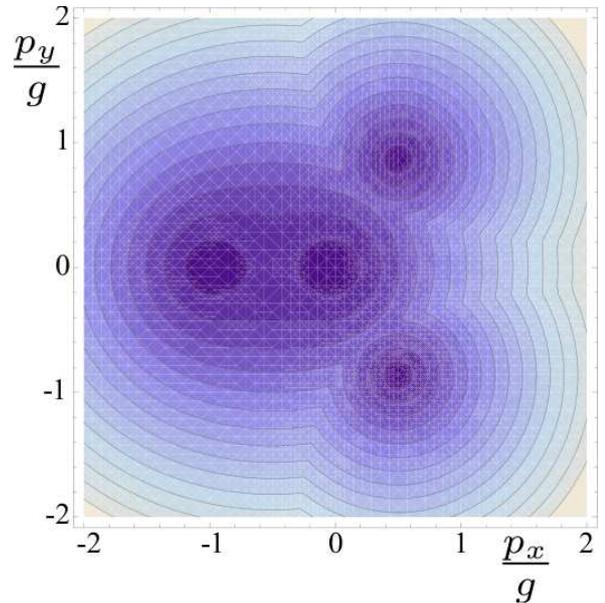}
\caption{\textbf{Fig.~6~}
Splitting of Fermi point with $N=4$ into four Fermi points with $N=1$ each in the spectrum of the Hamiltonian where in Eq.~\eqref{4x4a} the element proportional to the coupling $g_{24}$ has been changed to $\sigma^-$ instead of $\sigma^+$. All the (finite) couplings are assumed to have the same magnitude $g$.}
  \label{other_4x4_hamiltonian_spectrum_contour}
\end{figure}

In general the Fermi point with arbitrary $N$ may give rise to the spectrum
\begin{equation}
E^2=\gamma_N^2 p^{2N}\,.
 \label{NSpectrum}
\end{equation}
For graphene this spectrum is proposed in Ref. \cite{Castro2009}.
 The discussed symmetry of matrix elements $g_{mn}$ extended to  $2N\times 2N$ matrix gives 
\eqref{NSpectrum} with the prefactor
\begin{equation}
\gamma_N=\frac{1}{|g_{12}||g_{23}|\ldots |g_{N-1,N}|}\,.
 \label{NSpectrum2}
\end{equation}
Violation of this symmetry may lead to splitting of the multiple Fermi point  into
  $N$ elementary Fermi points -- Dirac points with $N=1$ and `relativistic' spectrum
  $E^2 \propto p^{2}$. For $N=4$ the splitting is shown in Fig. 
  \ref{other_4x4_hamiltonian_spectrum_contour}.

The effective Hamiltonian describing fermions in the vicinity of multiple Fermi point is
$H=\sigma^-p_+^N + \sigma^+p_-^N$, see  \cite{Manes2007}. An example of the effective Hamiltonian describing the multiple Fermi point with topological charge $N\equiv N_3$ in 3+1 systems is \cite{VolovikKonyshev1988}
\begin{equation}
H=\sigma_z p_z + \sigma^-p_+^N + \sigma^+p_-^N\,.
 \label{3+1}
\end{equation}
This Hamiltonian has the spectrum $E^2=p_z^2 + p_\perp^{2N}$,  which has linear dispersion in one direction and non-linear dispersion in the others.

\section{Conclusion}

 Using the momentum-space topology we found the classes of Hamiltonians which exhibit the nonlinear Dirac-like spectrum $E_\pm=\pm \gamma_N p^N$. Such spectrum arises, when there is a  symmetry which keeps together the Fermi points of different fermionic species, i.e. prohibits splitting into elementary Fermi points, i.e. Dirac points with $N=1$. Similar symmetries exist between the fermions in the Standard Model of particle physics, see \cite{Volovik2010} and references therein.
 
 There is a major difference between our approach and that of Kitaev who studied the classes of Hamiltonians  
 \cite{Kitaev2009}.  Instead of classification of Hamiltonians we make the topological classification of the ground states taking into account the symmetry of the ground state. We use the top-bottom procedure \cite{Hu2009}, i.e. consider from the very beginning the  microscopic interacting many-body system, and consider its properties at low energy. The objects of the topological classification are response functions in the background of the ground state, which belongs to some symmetry class. We use the single particle Green's function, but the higher order Green's function can also be appropriate.  The single particle Green's function matrix at zero energy represents the effective single-particle  Hamiltonian ${\cal H}({\bf p})=G^{-1}(\omega=0,{\bf p})$, which by definition is non-interacting.  In Refs. \cite{Fidkowski2010,Fidkowski2009,Pollmann2010} the opposite procedure is used: the authors start 
with  non-interacting Hamiltonians and extend them to the interacting Hamiltonians. This procedure makes sense when one constructs the artificial fermionic or bosonic systems. But it is not applicable to the natural many-body systems which are interacting on the microscopic level: such bottom-top procedure is not unique, and it is unable to restore the original microscopic Hamiltonian.  

Our approach is also applicable to the vacuum of particle physics. The Lorentz symmetry prohibits the splitting of the Dirac points, but it also prohibits the non-linear non-relativistic spectrum. Situation changes if the Lorentz symmetry is viewed as an emergent phenomenon, which arises near the Dirac point (Fermi points with $N=\pm 1$). In this case both variants are possible, depending on symmetry: splitting of Dirac points or formation of non-linear non-relativistic spectrum in the vicinity of the multiple Fermi point. In both cases the mixing of fermions violates the effective Lorentz symmetry in the low-energy corner. This phenomenon,  called the reentrant violation of special relativity \cite{Volovik2001}, has been discussed for $N_F = 3$ fermion families in relation to neutrino oscillations
\cite{Klinkhamer2006}. Influence of possible discrete flavor symmetries on neutrino mixing has been reviewed in Ref.
\cite{Altarelli2010}.

This work is supported in part by the Academy of Finland, Centers of excellence program 2006--2011, the European Research Council (Grant No. 240362-Heattronics), and the Khalatnikov--Starobinsky leading scientific school (Grant No. 4899.2008.2)

\end{document}